\documentclass[%
 reprint,
 amsmath,amssymb,
 aps,
]{revtex4-2}
\newcommand{\omg}{\omega}
\newcommand{\dg}{^\dagger}
\newcommand{\rmd}[1]{}
\rmd{fix remaining refs}
\usepackage{color}
\usepackage{textcomp}
\usepackage{amsmath}
\usepackage{graphicx}
\usepackage{float}
\newcommand{\od}[1]{^{(#1)}}
\newcommand{\rg}{\(^\text{\textregistered}\)}
\newcommand{\avg}[1]{\langle #1 \rangle}

\begin{document}
\author{Han Liu}
\email{qwerty.liu@mail.utoronto.ca}
\author{Meng Lon Iu}
\author{Noor Hamdash}
\author{Amr S. Helmy}
\email{a.helmy@utoronto.ca}
\affiliation{The Edward S. Rogers Sr. Department of Electrical \& Computer Engineering, University of Toronto}

\title{Towards Arbitrary Time-frequency Mode Squeezing with Self-conjugated Mode Squeezing in Fiber}

\begin{abstract}
	Optical parametric amplification generates squeezed light in device-specific sets of time–frequency eigenmodes, and it has been widely accepted that detection and utilization of squeezing must comply with this modal constraint. We show that this constraint can be considerably relaxed under the continuous-wave pump and broadband phase-matching approximation, where the modal decomposition is non-unique. Specifically, any time-frequency mode with “self-conjugated” spectral symmetry can approximate a squeezing eigenmode, and partial homodyne detection can herald squeezing in arbitrary time-frequency modes. We demonstrate this using a high-efficiency, low-loss all-fiber source, measuring \(4.38 \pm 0.11\;\text{dB}\) and \(0.88 \pm 0.09\;\text{dB}\) squeezing on partially coherent and chaotic self-conjugated modes. Using a bichromatic self-conjugated mode with reduced local-oscillator noise, we achieve \(7.50 \pm 0.12\;\text{dB}\) squeezing, which represents the highest level reported for fully guided-wave squeezing sources based on \(\chi^{(2)}\) and \(\chi^{(3)}\) nonlinearities.
\end{abstract}

\maketitle
\section{Introduction}
Squeezed states of light are quantum states that exhibit reduced uncertainty in one quadrature of the electromagnetic field, enabling noise levels below the shot-noise limit of classical light sources. This property has proven essential in advancing a broad range of scientific and engineering applications, including quantum communication~\cite{slusher1990squeezed,li2002quantum,gottesman2003secure,eberle2013gaussian,fanizza2021squeezing}, quantum computing~\cite{gottesman2001encoding,filip2005measurement,yoshikawa2007demonstration,asavanant2019generation,vernon2019scalable}, and quantum metrology~\cite{abbott2016observation,lawrie2019quantum,polzik1992spectroscopy,michael2019squeezing,casacio2021quantum,haus1991fiber,eberle2010quantum,grace2020quantum,xu2019sensing,sudbeck2020demonstration,yap2020generation,zhang2021distributed}. The critical importance of high-quality squeezed light sources in these areas motivates the development of sources with high squeezing, diverse functionalities and compact form factors\cite{yamamoto1992photon,fox1995squeezed,dutt2015chip,zhang2021squeezed}.

Since the initial experimental demonstration of squeezed vacuum noise \cite{slusher1985observation}, extensive research has been devoted to innovating squeezed light sources to achieve significant levels of shot noise reduction \cite{andersen201630}. The current pinnacle in squeezing (15 dB) \cite{vahlbruch2016detection} is achieved through free-space optical parametric amplification (OPA). Guided wave platforms \cite{kaiser2016fully,takanashi20204,vaidya2020broadband,inoue2023toward}, though characterized by lower squeezing, offer distinct advantages in terms of robustness and ease of integration, owing to their compact form factor and resilience to environmental perturbations. Integrated optical waveguides \cite{serkland1995squeezing,pysher2009broadband,mondain2019chip,kashiwazaki2020continuous,kashiwazaki2023over} and optical fibers \cite{shirasaki1990squeezing,rosenbluh1991squeezed,bergman1991squeezing,drummond1993quantum,bergman1994squeezing,friberg1996observation,schmitt1998photon,spalter1998observation,krylov1998amplitude,fujiwara2009generation,guo2016generation,kalinin20247} emerge as the two principal platforms for squeezed light generation, each presenting unique challenges and advantages.\\

Beyond the magnitude of squeezing, the time-frequency properties of squeezed light sources are of formidable significance for the utilization of squeezing. This is not only because time-frequency modes are ideal high-capacity carriers of quantum information \cite{mower2013high,liu2020joint,lu2023frequency,seshadri2022nonlocal,blakey2022quantum}, but also because optical protocols often have platform-specific requirements for light sources with different time-frequency properties. For instance, low-coherence light is required for white light interferometry \cite{diddams1996dispersion,wyant2002white} and fiber-optical gyroscopes \cite{lefevre2022fiber}, while temporally multiplexed two-mode squeezed states can be used to generate 2D cluster states for quantum computing \cite{larsen2019deterministic,asavanant2019generation}. High-speed optical communication and information processing protocols, such as quantum frequency conversion \cite{eckstein2011quantum,brecht2015photon,kumar1990quantum}, may also benefit from squeezed light with protocol-dependent time-frequency characteristics. Additionally, certain protocols require dynamically tunable sources, such as spectroscopy and frequency-domain optical coherence tomography that use frequency-sweeping sources.\\

Given these diverse application needs, a versatile squeezed light source with flexible and dynamically controllable time-frequency characteristics would be highly beneficial. However, implementing such a modality is encumbered by the fundamental constraints of the underlying OPA processes: OPA generates squeezing in a specific set of intrinsic time-frequency modes, depending on dispersion and pump conditions \cite{braunstein2005squeezing,lvovsky2015squeezed,wasilewski2006pulsed,de2006multimode,roslund2014wavelength}. This modal structure can be controlled through dispersion engineering and shaping the pump pulses\cite{ansari2018tailoring,arzani2018versatile}, which is a useful technique for generating spectrally separable states \cite{mosley2008heralded,eckstein2011highly}. Nevertheless, the achievable tunability is constrained by the extent to which the material dispersion and pump light properties can be manipulated and affect the OPA modal structure. Mechanical tuning of the optical cavity also provides some degree of freedom to tune the squeezing process but with limited tuning speed and range depending on the physical implementation \cite{hagemann202410}. Without changing the modal structure of the squeezed light source, dynamic manipulation of quantum light after its generation can also be done through temporal imaging\cite{karpinski2017bandwidth,foster2009ultrafast}, but at the cost of insertion loss that may degrade measurable squeezing.
\begin{figure*}[ht]
	\centering
	\includegraphics[width=2\columnwidth]{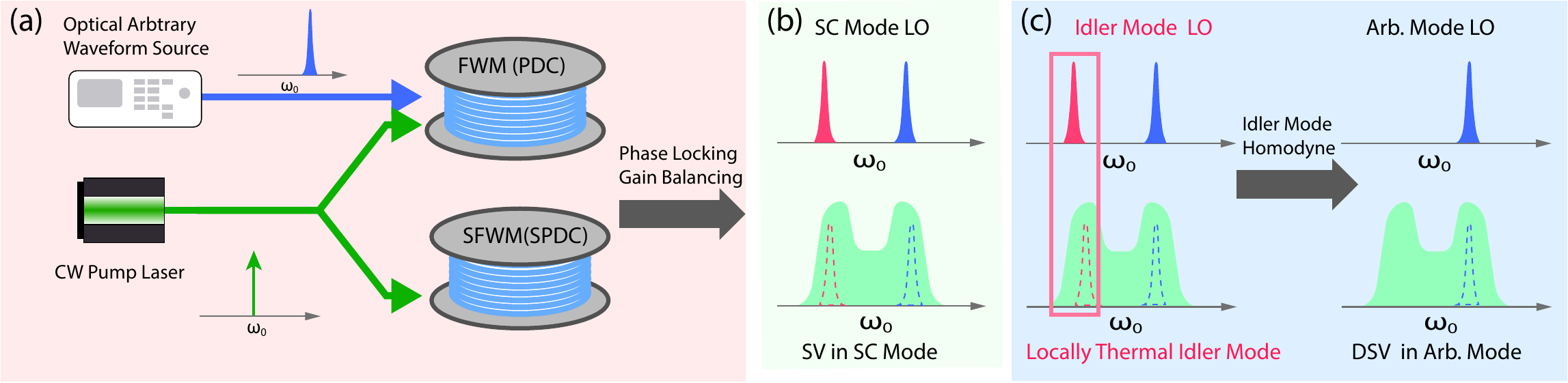}\\
	\includegraphics[width=1.8\columnwidth]{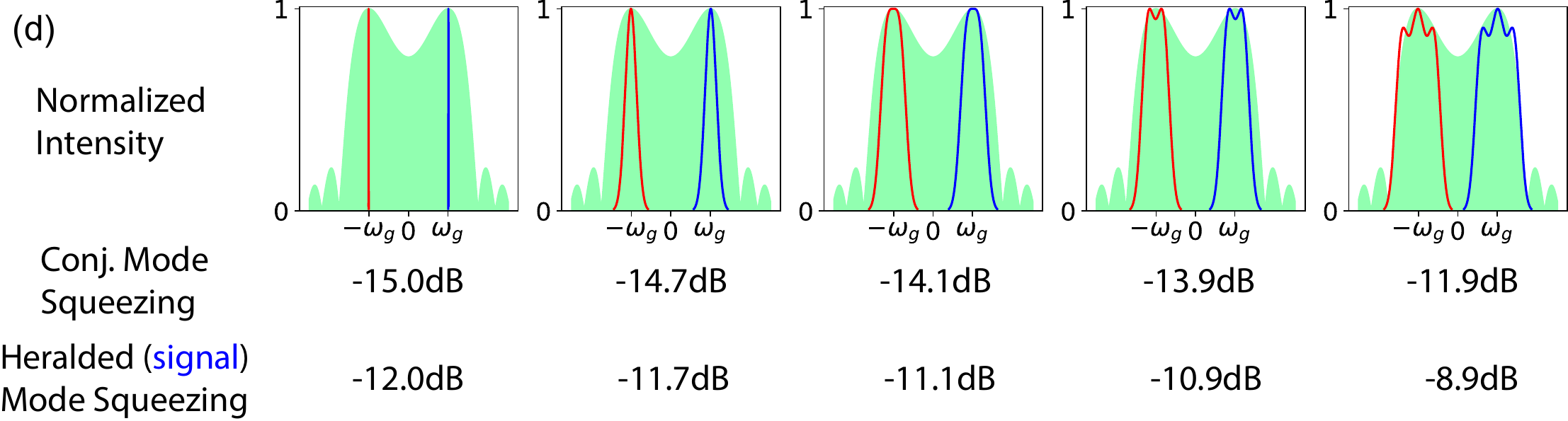}
	\caption{(a) Generation of SC mode squeezing through SFWM or spontaneous parametric down-conversion (SPDC). The corresponding LO light can be generated through FWM or parametric down-conversion process (PDC), respectively. (b) The spectral intensity of LO light (solid fill), squeezed vacuum (SV) on the SC mode (dashed curve), and the frequency-dependent squeezing parameter \(|\xi(\omg)|\) (green background). Red and green colors denote the signal and idler frequency branches, respectively. (c) The procedure of converting squeezing on the mode into a displaced squeezed vacuum (DSV) on the signal mode through partial homodyne detection on the idler mode. (d) Examples of different signal seed light spectrum (red curve) and its corresponding idler mode spectrum (blue curve) as well as the corresponding SC mode squeezing and heralded signal mode squeezing (for 1.5 km single mode fiber as the nonlinear medium and the peak gain detuning is \(2\pi\times50.62\) GHz from the pump frequency). Both SC mode squeezing and heralded squeezing decrease as the spectral amplitude of the SC mode extend beyond the spectral range where \(|\xi(\omg)|\) is close to its maximum.  \label{illustrate_conj_herald}}
\end{figure*}

A novel approach to tailoring quantum light properties in a dynamic fashion is to leverage the entanglement of light between two distinct channels: measurement on one (heralding) channel will herald light in the other (heralded) channel into a specific quantum state that depends on the measurement basis and result. This concept has been successfully applied in generating arbitrarily shaped (in the time-frequency domain) single-photon states using time-frequency entangled photon pairs \cite{franson1992nonlocal,bellini2003nonlocal,sych2017generic,ansari2018heralded,averchenko2020efficient}. Additionally, a similar concept is applied in measuring spatially multimode squeezing. In \cite{embrey2015observation}, the local oscillator for spatial homodyne detection of squeezing consists of a pair of local oscillator (LO) beams that satisfy certain spatial and phase conjugation relations. Each of the conjugated LO beams is free to take an arbitrary (but correlated) spatial profile. This degree of freedom allows observation of squeezing in arbitrarily shaped spatial modes, which is crucial for imaging and microscopy with quantum-enhanced spatial resolution \cite{kolobov2000quantum}. In principle, the concept of squeezed mode shaping can be extended into the time-frequency domain, which features significantly higher capacity and scalability of dimension. Such a squeezing source, if implemented, has the potential to benefit a wide range of practical and scientific applications that rely on squeezed light.\\

Drawing inspiration from the aforementioned previous works, we report here a novel technique of generating squeezed light in self-conjugated time-frequency modes: modes that have spectral intensity and phase that are symmetric and anti-symmetric around some center frequency. The key observation is that any self-conjugated mode can be considered as an eigenmode of an OPA based squeezed light source, in the limit of broadband phase-matching and continuous-wave (cw) pump. It is further shown that squeezing on self-conjugated modes also paves the way towards dynamically reconfigurable squeezed light sources: projective (homodyne) measurement on one branch of the self-conjugated mode can herald squeezing in the other branch. Since the only requirement between the two branches is self-conjugation, the heralded squeezing in one branch is allowed to have arbitrary temporal and spectral amplitude. We term such a capability as arbitrary time-frequency mode squeezing.\\ 
To experimentally verify the validity of this approximation, an all-fiber squeezing source is constructed to measure squeezing on self-conjugated modes with randomized time-frequency properties, i.e., partially coherent modes defined by noise-modulated laser light and chaotic time-frequency modes defined by amplified spontaneous emission (ASE) light \cite{liu2023compact}.
The time-frequency domain flexibility of squeezed light in self-conjugated modes also provides practical advantages in observing high squeezing with a simple optical setup. In particular, the squeezing source and detection system are implemented entirely with standard off-the-shelf fiber-optical components that feature exceptionally low propagation and coupling loss. The problem of guided acoustic wave Brillouin scattering (GAWBS) that is present in fiber squeezing sources based on self-phase modulation is completely avoided because the squeezed time-frequency mode can be tuned away from the GAWBS noise band.\\

The rest of this paper is organized as follows. We first introduce the concept of self-conjugated mode squeezing in a fiber OPA and show its relation to generating squeezed light in arbitrary time-frequency modes. Then, we experimentally demonstrate generating and measuring squeezed light in three different self-conjugated modes with different characteristics. We also theoretically show how self-conjugated mode squeezing can lead to arbitrary time-frequency mode squeezing under the broadband phase-matching approximation. We conclude by discussing paths to performance enhancement potential of applications in practical squeezing-based protocols.

\begin{figure*}[ht]
	\centering
	\includegraphics[width=2\columnwidth]{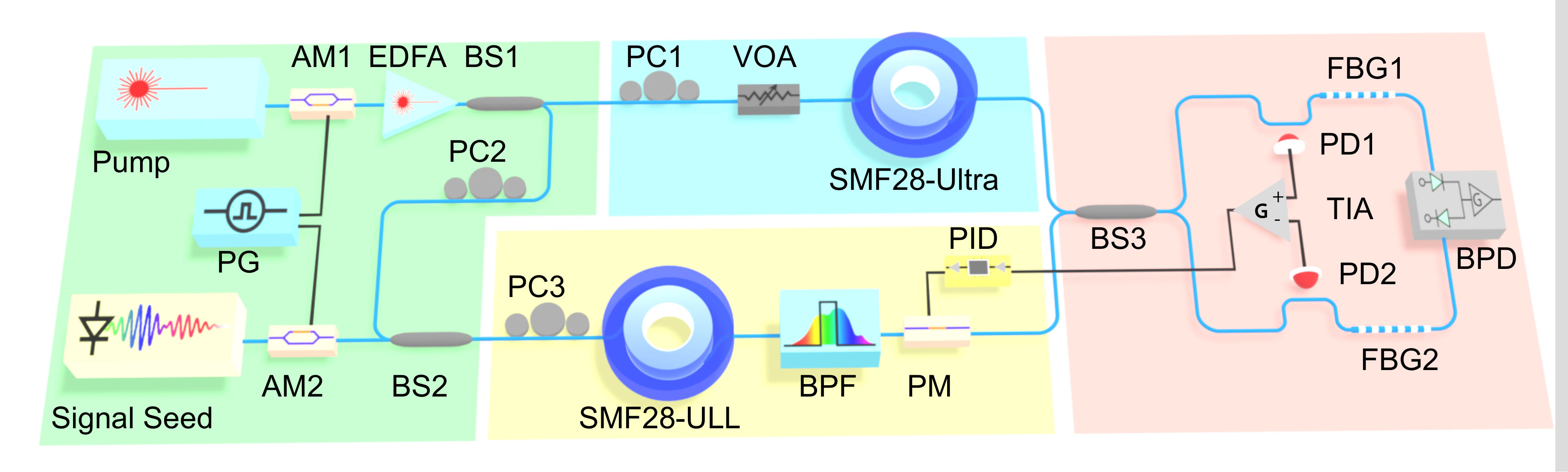}
	\caption{The schematic of the squeezed light system setup consists of four different sections (green: the pump and signal seed light source, blue: the SFWM arm, yellow: the FWM arm, red: the BHD section). PG: pulse generator; AM1-2: electro-optical amplitude modulators; EDFA: erbium-doped fiber amplifier; PC1-3: inline fiber polarization controllers; BS1: a 50:50 fiber coupler for distributing pump light into the SFWM and FWM arms; BS2: a 10:90 fiber coupler that mixes pump and signal seed light before FWM; VOA: variable optical attenuator; BPF: tunable band-pass filter; PM: electro-optical phase modulator; PID: proportional-integral-derivative controller with peripheral circuits; BS3: the BHD coupler; FBG1-2: fiber Bragg gratings for pump light filtering; PD1-2: inline optical power monitors; TIA: transimpedance amplifier; BPD: balanced photodetector pair.\label{experimental_setup}}
\end{figure*}

\section{Theoretical Formalism}
In this section, we begin by providing a theoretical framework for defining self-conjugated (SC) time-frequency modes and describe their evolution in an OPA process. Then we show how squeezing on an SC mode can be converted into heralded squeezing on an arbitrary mode with partial homodyne detection.
\subsection{Generation of Squeezing on Self-Conjugated Modes}
	We consider spontaneous four-wave mixing (SFWM) as the OPA process (with vacuum input) for a single spatial-polarization mode fiber pumped by a laser at frequency \(\omg_0\) as illustrated in Fig.\ref{illustrate_conj_herald} (a-b). Furthermore, we assume that the pump laser is cw or the squeezed field of interest temporally overlaps within the constant-power portion of the pump light pulses. Then the SFWM evolution \(U\) can be described by \cite{suppmat}:
\begin{gather}
	U  = \exp\left(-\int_{\omg_0}^{+\infty}(\xi(\omg)a\dg(\omg)a\dg(2\omg_0-\omg)-h.c.)d\omg\right)\label{USFWM0}
\end{gather}
where \(a(\omg)\) is the frequency domain annihilation operator of the fiber mode and \(h.c.\) stand for Hermitian conjugate. We customarily designate the frequency range higher (lower) than \(\omg_0\) as the signal (idler) branch, respectively. The SFWM phase-matching condition is characterized by the joint spectral amplitude \(\xi(\omg)=\xi(2\omg_0-\omg)=|\xi(\omg)|\exp(i\phi(\omg))\). We term the spectral range of non-negligible \(|\xi(\omg)|\) as the phase-matching spectrum.\\

Considerable insight can be obtained by taking the broadband phase-matching limit, i.e. \(\xi(\omg)=|\xi(\omg)|= r\) is a constant real number for different \(\omg\) (the general case is analyzed in the Methods section). Then \(U\) admits a non-unique decomposition:
\begin{gather}
	U  = \exp\left(-(r \sum\limits_n A_n\dg B_n\dg -h.c.)\right)\\
	A_n = \int_{\omg_0}^{+\infty} f^*_n(\omg) a(\omg)\hspace{0.2cm} B_n =\int_{-\infty}^{\omg_0} f_n(2\omg_0-\omg) a(\omg)\label{ABTMSQ}
\end{gather}
where \(f_n(\omg)\) is an arbitrary set of orthonormal function for \(\omg>\omg_0\). Since the first function \(f_0(\omg)\) can be arbitrarily specified (and the rest of \(f_n(\omg)\) can be obtained through the Gram-Schmidt procedure), Eq.\eqref{ABTMSQ} essentially state that an arbitrary pair of signal mode \(A_0\) and idler mode \(B_0\)  are two modes squeezed by \(U\), as long as their complex spectral amplitude  are mutually symmetric in magnitude and anti-symmetric in phase (e.g., \(f_0(\omg)\) and \(f_0^*(2\omg_0-\omg)\)).\\

The two-mode squeezing of arbitrary signal and idler mode pairs can also be converted into single-mode squeezing. 
Consider a supermode \(C\) that is obtained by symmetrically mixing a signal and idler mode pair \(A_0\) and \(B_0\), then \(C\) is single mode squeezed with squeezing parameter \(-r\):
\begin{gather}
	U\dg CU = \cosh(r)C-\sinh(r)C\dg\\
	C =\frac{A_0+B_0}{\sqrt{2}} = \int_{-\infty}^{+\infty} f^*_\text{conj}(\omg) a(\omg) d\omg\\
	f_\text{conj}(\omg) = 
	\begin{cases}
		\frac{1}{\sqrt{2}}f_0^*(2\omg_0-\omg) &(\omg\le \omg_0)\\
		\frac{1}{\sqrt{2}}f_0(\omg) &(\omg> \omg_0)\\
	\end{cases}\label{SCDEF}
\end{gather}
By the arbitrarity of \(f_0(\omg)\), any time-frequency mode satisfying the SC symmetry of Eq.\eqref{SCDEF} (symmetric in intensity and anti-symmetric in phase around \(\omg_0\)) can be regarded as the mode \(C\). We term such modes \(C\) as SC modes. When the joint spectral amplitude \(\xi(\omg)\) is not real, the above argument still applies but with a modified definition of SC modes that takes into account the dispersion of \(\xi(\omg)\)\cite{suppmat}. When the broadband phase-matching approximation is not perfectly satisfied (i.e. \(|\xi(\omg)|\ne r\) for some \(\omg\)), the squeezing for the SC mode is given by the average of \(|\xi(\omg)|\) weighted over \(|f(\omg)|^2\) (see the Methods section).\\

To measure squeezing on some SC mode \(C\) through balanced homodyne detection (BHD), LO light with SC spectral amplitude \(f_\text{conj}(\omg)\) needs to be prepared. Such LO light can be derived from the output of another four-wave mixing (FWM) channel whose pump is coherent with that in SFWM. The FWM process also needs to be seeded by some \textit{signal seed light} with an arbitrary spectral amplitude \(f_0(\omg)\) that lies within the signal (or idler) branch, as illustrated in the top half of Fig.\ref{illustrate_conj_herald}(a-b). The FWM output consists of both a amplified version of the signal seed light (signal light) and its spectral mirror image in the idler branch (idler light): 
\begin{gather}
	f_\text{FWM}(\omg) = \sqrt{G}f_0(\omg) + \sqrt{G-1} f_0^*(2\omg_0-\omg)\label{need_for_balancing}
\end{gather}
It can be seen that, after selective attenuation of the signal light by a factor \((G-1)/G\), the spectral ampliutude of the signal and idler light  is proportional to some SC spectral amplitude \(f_\text{conj}(\omg)\). Therefore this combination of signal and idler light can serve as the LO for the corresponding SC mode. This technique of spectral tailoring the LO light is different from linear or nonlinear pulse shaping technique reported previously\cite{aytur1992squeezed,eto2008observation} because it is the mutual correlation between the signal and idler light that is shaped instead of the temporal or spectral property that is local to either the signal or the idler light. 

\subsection{Heralded Squeezing on Arbitrary Signal Modes}
Generation of arbitrary SC mode squeezing can serve as an intermediate step towards generating squeezing on time-frequency modes that take arbitrary temporal or spectral amplitude (as illustrated in Fig.\ref{illustrate_conj_herald} (c)): a partial homodyne detection on some arbitrary idler mode \(B_0\) will herald the corresponding signal mode \(A_0\) into a displaced squeezed vacuum state\cite{suppmat}. This can be done, for example, by spectrally separating the idler branch from the signal branch and performing independent BHDs. 
If the broadband phase-matching approximation is satisfied, the heralded state on \(A_0\) is a displaced squeezed vacuum state with coherent component \(\avg{X_\text{heralded}}\) and squeezing parameter \(r_\text{heralded}\):
\begin{gather}
	\avg{X_\text{heralded}} = -\text{tanh}(r)\gamma\label{amplitude_and_gamma}\\
	r_\text{heralded}= -\frac{1}{2}\log(\text{cosh}(2r)) \label{pure_heralded_squeezing}
\end{gather}
where \(\gamma\) is the outcome of the heralding BHD on the idler mode. When the idler mode (\(B_0\)) measurement result is regarded as unknown, the photon statistics of heralded signal mode are reduced to usual thermal statistics.  Figure \ref{illustrate_conj_herald}(d) shows examples of different signal mode spectral amplitude \(f_0(\omg)\), showing the dependence of SC mode squeezing and heralded squeezing on the overlap between \(|\xi(\omg)|\)  and \(f_0(\omg)\).\\ 

	The capability of heralding squeezing on an arbitrary time-frequency mode is an entanglement effect: different choices of \(A_0, B_0\) correspond to measuring correlation in locally incompatible bases that would otherwise violate the Heisenberg uncertainty principle without entanglement. This is closely related to shaping single photons using time-frequency entanglement and shaping spatial squeezing using spatial entanglement. Nevertheless, unlike indeterministic shaping of single photons, the generation of squeezing Eq.\eqref{pure_heralded_squeezing} is deterministic. The additional indeterministic coherent component Eq.\eqref{amplitude_and_gamma} can either be subtracted through linear interference or compensated in data processing after the heralded state is measured via BHD. The generation rate of heralded squeezing is limited by the electrical bandwidth of BHD, the repetition rate of pump pulses, or the phase-matching bandwidth, whichever is the lowest. This entanglement-based squeezing shaping technique offers advantages in terms of time and frequency domain flexibility over techniques that are based on pump pulse shaping and device engineering \cite{ansari2018tailoring,arzani2018versatile}. Nevertheless, it is at the expense of not only increased system complexity (heralding BHD) but also the heralding penalty of squeezing: Eq.\eqref{pure_heralded_squeezing} shows that in the limit of strong squeezing \(|\xi(\omg)|\gg 1\), heralded squeezing is 3 dB less than the squeezing directly measured on the SC mode. This is because the heralding measurement on the idler mode reveals a portion of the information of the joint quantum state of the signal and idler mode.

\rmd{phase-matching or phase-matching}

\section{Experimental Setup and Squeezing Measurement} 
\subsection{Experimental Setup}
The setup schematics are shown in Fig. \ref{experimental_setup} (see the Methods section for implementation details). In the pump source section, nanosecond-wide pump pulses with uniform peak power are generated for both the SFWM and FWM arms. It is obtained by amplitude modulating a cw laser with pulsed waveform and then amplified with an erbium-doped fiber amplifier (EDFA). To block the broadband ASE from the EDFA that is inband with the squeezed field of interest, the pump light is bandpass-filtered with a combination of a fiber Bragg grating and a circulator. The reason for using pulsed as opposed to cw pump light is to achieve efficient nonlinear interaction (high peak power) and suppress stimulated Brillouin scattering\cite{hansryd2002fiber,agrawal2000nonlinear}. The SFWM process takes place inside a 1.5 km spool of Corning\rg SMF28-ultra fiber spool (room temperature, without cooling) whose output is directly spliced to the BHD section. The FWM process takes place inside another 1.5 km long Corning\rg SMF28-ULL fiber fiber spool, taking both pump light and signal seed light as the input. The outcome of the FWM consists of conjugated signal and idler light with different power (see Eq.\eqref{need_for_balancing}). Such difference in power is manually balanced by inserting a tunable bandpass filter on the FWM output and tuning its cut-on frequency such that a slight attenuation is applied to the signal light. The tunable bandpass filter also helps block unwanted spectral components (due to cascaded FWM effects) of the LO. After balancing, the signal and idler light in conjunction serve as the LO for BHD. The BHD section consists of a 50:50 BHD coupler through which the LO interferes with the SFWM output and a balanced photodiode pair (InGaAs, 95\% quantum efficiency, including coupling losses from collimation optics). The output of the photodiode pair is amplified with a transimpedance amplifier and analyzed by an electrical spectrum analyzer (ESA). 

\subsection{Phase Stabilization}
To observe strong squeezing with temporal stability, the relative phase between the LO and squeezed vacuum needs to be locked with an optical phase-locked loop. This is done by using the interference of pump light (from the SFWM and the LO arm on the BHD coupler) as an error signal to control the LO phase via a phase modulator in the LO arm. The pump power interference is measured with two low-loss inline optical power monitors before the two pump-blocking FBGs. The obtained photocurrent is amplified and processed by a high-speed proportional-integral-derivative (PID) controller that is implemented with analog-to-digital converters and a field programmable gate array. The output of the PID controller is digital-to-analog converted and amplified to drive the phase modulator. The locked relative phase is digitally tuned by adjusting the PID input offset to maximize measurable squeezing. To ensure that backreflected pump light from the two fiber Bragg gratings (FBGs) does not re-enter the SFWM arm, the relative phase between the backreflection of two FBGs are controlled by a fiber stretcher in one of the BHD coupler output arms.\\
\subsection{Performance Benchmarking}
\begin{figure}
	\includegraphics[width=0.8\columnwidth]{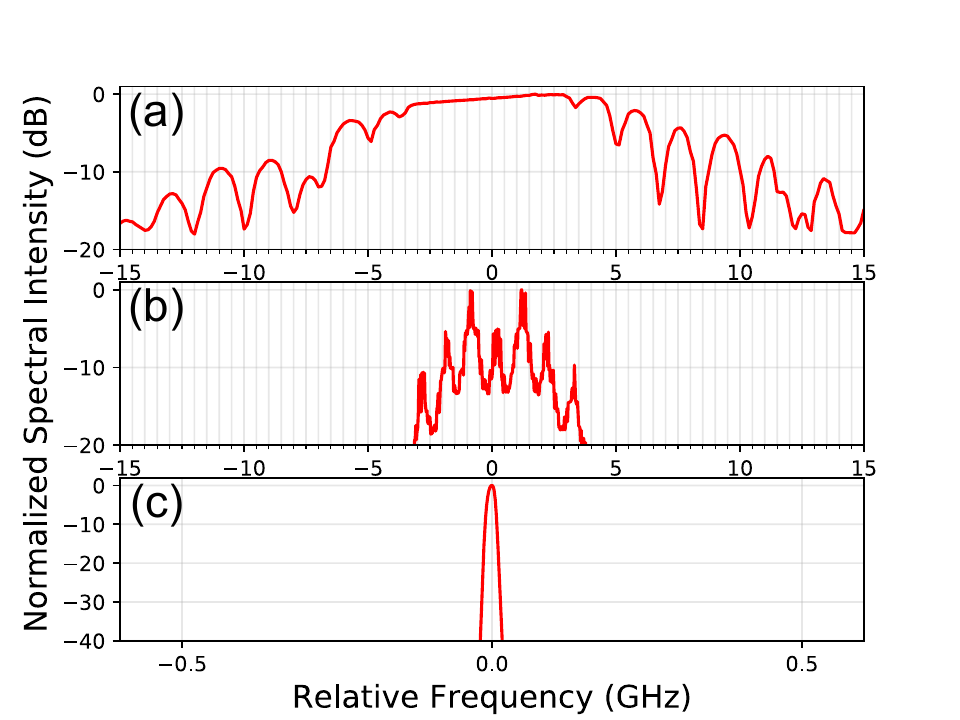}
	\caption{The spectra of (a) filtered ASE light and (b) noise-modulated laser (with 1 GHz sine modulation) light and (c)coherent laser light. The frequency axes are normalized relative to their center frequencies (1549.55 nm, 1549.65 nm, and 1549.65 nm), respectively. The spacing of frequency and amplitude grid lines is equal for all three plots\label{spectra}.}
\end{figure}

To quantify the capability to generate and measure squeezing on arbitrary SC modes, the signal seed light of FWM is switched between three different sources to generate LO light on different SC modes. The first signal seed source that serves as a performance baseline is a single frequency laser at 1549.65 nm. The second source is the same laser but going through three stages of phase modulation: (1) 40 MHz laser coherence control and (2) low frequency modulation with 300 MHz random waveform and (3) high frequency modulation with 0.5-1 GHz sine signal. The third source is filtered ASE light obtained by cascading narrowband optical filters and fiber amplifiers after a broadband ASE source. The spectra of different signal seed light sources are shown in Fig.\ref{spectra}(a-c) respectively, measured through balanced heterodyne with a frequency swept source. It is worth noting that unlike previously reported squeezing measurements with incoherent pump \cite{kumar1990squeezed}, the noise-modulated and chaotic signal seed light used here is completely independent and non-phase-locked with the SFWM pump light. All signal seed light sources undergo pulse amplitude modulation (3 ns) to fit within the temporal span of the pump pulse. To benchmark the squeezing performance, the squeezed shot noise level of the homodyne output is measured with the ESA at specific frequencies when the SFWM pump light is unblocked. When the SFWM pump light is blocked, the BHD output level is defined as the apparent shot noise level that may include the relative intensity noise of the LO light. For reference, the true shot noise level of a narrow linewidth laser at the same power as the LO is also measured. 

\begin{figure}[h]
\centering
\includegraphics[width=0.7\columnwidth]{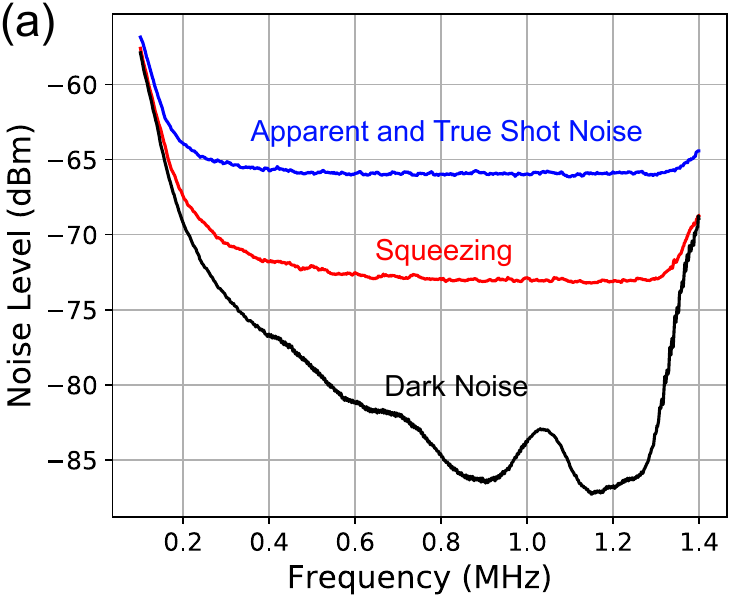}\\
\includegraphics[width=0.7\columnwidth]{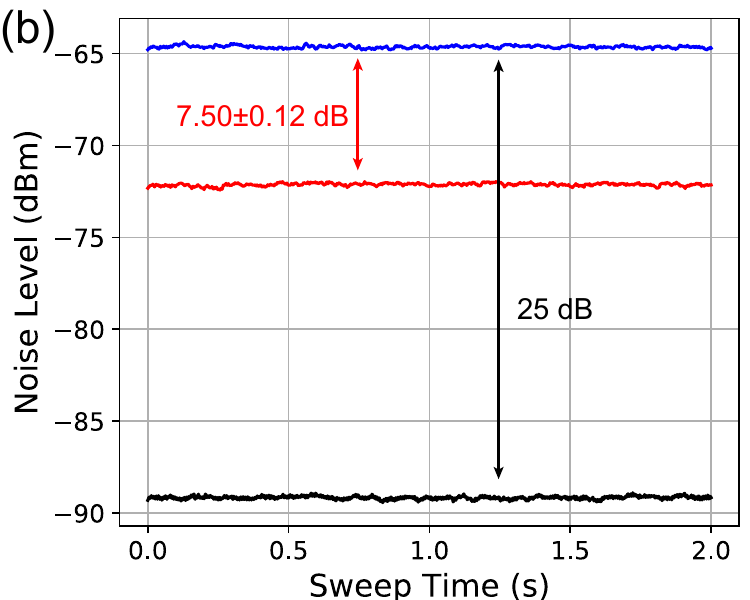}\\
\includegraphics[width=0.7\columnwidth]{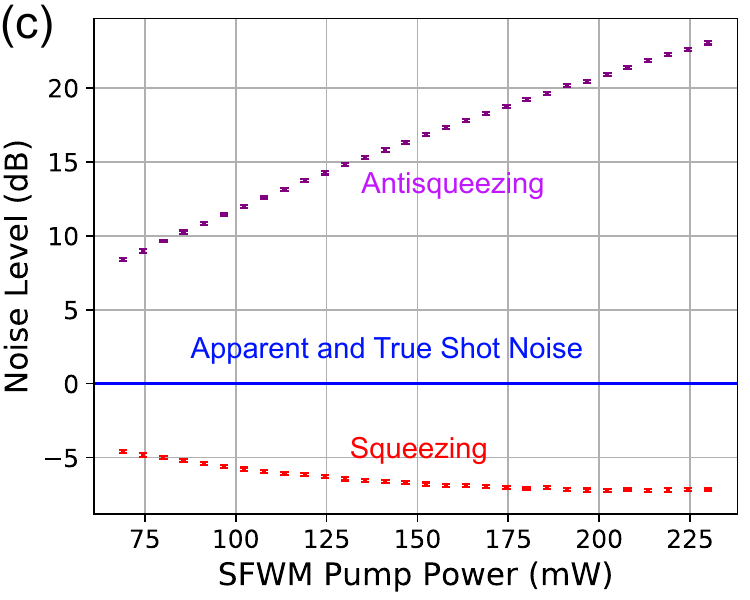}\\
\caption{The squeezed noise level (red), shot noise level (blue), and dark noise level (black) as a function of (a) electrical frequency and (b) time (zero span at 1.05 MHz) when coherent laser light is used as signal seed light. (c) the dependence of maximal squeezing (red) and antisqueezing (purple) relative to the shot noise level and their uncertainties as a function of SFWM pump average power.\label{laser_squeezing} }
\end{figure}

With the single-frequency laser as the signal seed light, the generated LO light consists of two discrete frequency components. This is similar to a bichromatic LO light\cite{marino2007bichromatic,embrey2016bichromatic,yang2021squeezed,shi2020observation} except that the signal and idler light of the LO do not have to be phase-locked with the pump light. The measured apparent shot noise level is confirmed to be the same as the true shot noise level. The linearity of the detector is also confirmed by measuring the shot noise level for different LO powers. The maximal squeezing measured (Fig.\ref{laser_squeezing}(a,b)) is \(7.50\pm0.12\) dB at 1.05 MHz. The dependence of squeezing and antisqueezing on the pump average power is shown in Fig.\ref{laser_squeezing}(c). The maximal squeezing peaks at around 219 mW when measured antisqueezing reaches 22 dB. The loss of the SFWM arm includes 7.2\% fiber propagation loss, 2.2\% fiber Bragg grating transmission loss, 0.2\% power monitor loss, 1.5\% fiber coupler loss, and 5.0\% detector inefficiency (including the fiber-to-diode coupling loss). Taking into account the -25 dB detector dark noise level, the estimated internal squeezing at the SFWM output is around 10.2 dB. \\

To investigate the difference between the 22 dB antisqueezing and the 10.2 dB estimated squeezing at the SFWM output, we measure the power spectrum of the SFWM output through balanced heterodyne detection and compare it with the measured phase-matching spectrum \(|\xi(\omg)|\) (See the Methods section for measurement details). It is shown that in addition to the power of squeezed vacuum, excess quantum noise exists beyond the SFWM phase-matching spectrum. Such noise is confirmed to be temporally non-uniform (overlapping with the pump pulses) and polarization sensitive. We attribute this to the broadband spontaneous Raman scattering noises as predicted and analyzed in \cite{shapiro1995raman,voss2006raman} and experimentally witnessed in \cite{dong2008experimental}.
The previous quantum mechanical calculation \cite{shapiro1995raman,voss2006raman} of the impact of Raman noise on squeezing does not include the effect of loss, which is non-negligible (7.2\%) for our 1.5 km fiber. In the supplementary material, we present a simple analysis of the compound effect of loss and Raman scattering, with the assumption that the introduction of Raman noise can be modeled as linear mixing with an external noise reservoir that is uniform across the fiber length. Simulation results show that for parametric gain that corresponds to 22 dB antisqueezing, squeezing at the SFWM arm output is expected to be 10.3 dB, which is in close agreement with the experimental result.

\begin{figure}[h]
\centering
\includegraphics[width=0.7\columnwidth]{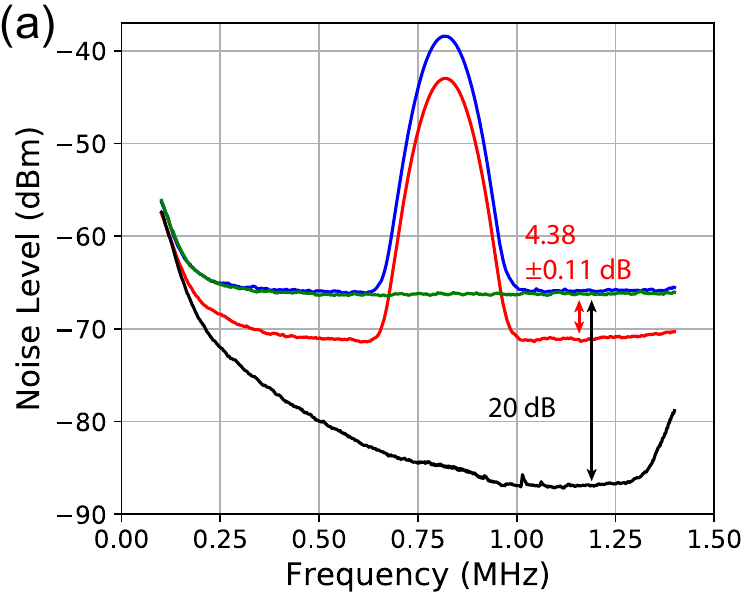}\\
\includegraphics[width=0.7\columnwidth]{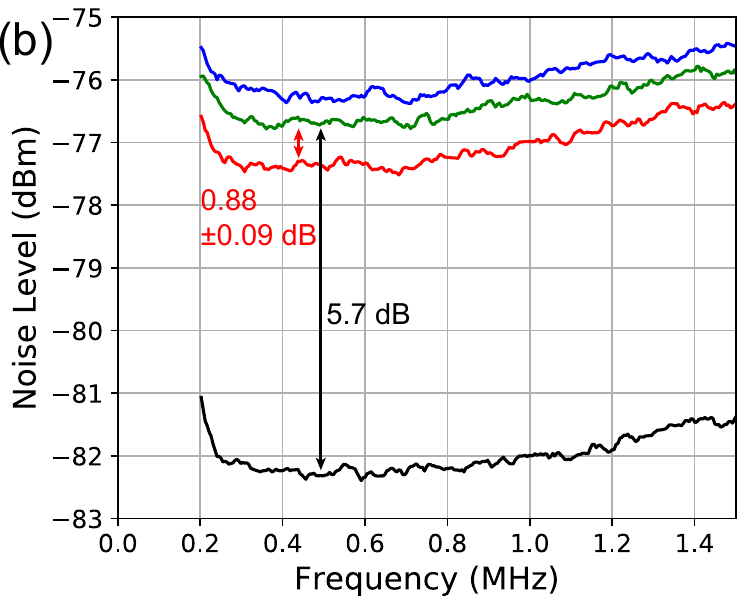}\\
\includegraphics[width=0.7\columnwidth]{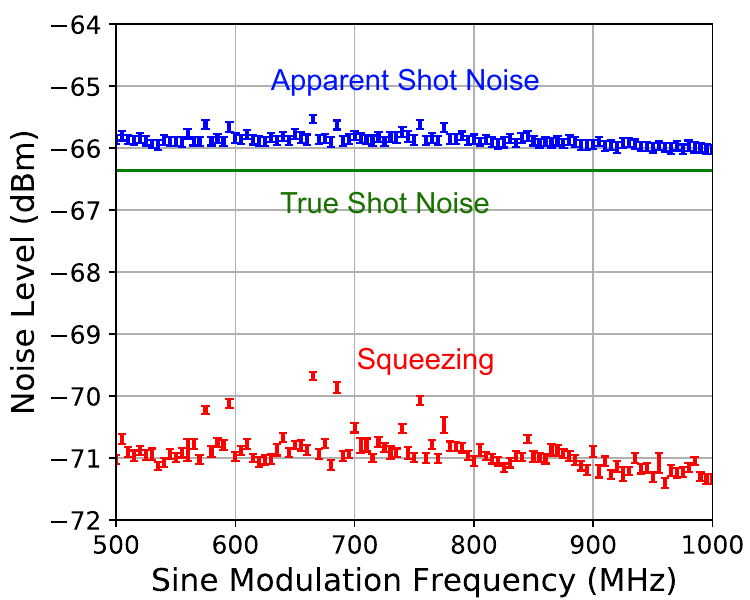}
\caption{The spectra of squeezed noise level (red), apparent shot noise level (blue), true shot noise level (green), and dark noise level (black) with (a) noise-modulated laser (with 1 GHz sine wave frequency modulation) and (b) filtered ASE light are used as the signal seed light. The labeled value of squeezing in (a) and (b) is defined as the difference between the squeezed noise level and the true shot noise level, both measured in separate zero-span ESA measurements at the specified frequencies. (c) the dependency of measured squeezing (at 1.15 MHz) on the sine modulation frequency. Blue and Red error bars represent the measured apparent shot noise and squeezing noise level. The green horizontal line represents the true shot noise level.
\label{chaotic_squeezing}}
\end{figure}

When a noise-modulated laser is used as the signal seed light, the measured squeezing at 1.15 MHz is reduced to 4.38 \(\pm\) 0.11 dB  (with 1 GHz sine modulation frequency and 219mW pump power, Fig. \ref{chaotic_squeezing} (a)) and is weakly dependent on the sine modulation frequency (Fig.\ref{chaotic_squeezing}(c)), indicating that the noise-modulated laser spectrum stays within the SFWM phase-matching bandwidth. The major reason that contributed to such reduction of squeezing as compared to without using modulation (7.50 dB) is because of the parasitic amplitude noise introduced by non-ideal phase modulation, this can be seen from the increase of shot noise level (0.74 dB) compared to that of a coherent laser of the same power. Taking such non-squeezable classical noise as well as the increase of dark noise (-20 dB below shot noise) into account, the measurable squeezing is predicted to be 4.37 dB, which is in close agreement with the experimental result. Filtered ASE light as the signal seed light features even higher relative intensity noise than noise-modulated laser light. To reduce the impact of such noise on squeezing measurement, a commercial balanced detector pair (Femto\rg HBPR 100M) with better common mode rejection (55dB) but lower efficiency (rated by 72\%) and higher detector dark noise level (-5.7 dB below the shot noise level) is used. Taking into account the residue relative intensity noise after common mode rejection (0.11 dB above shot noise), reduced quantum efficiency, and the increased dark noise, the predicted squeezing is 1.47 dB. Experimentally, 0.88\(\pm\)0.09 dB (at 490 kHz with \(\simeq\)50 mW pump power) of squeezing is measured (Fig.\ref{chaotic_squeezing} (b)). This discrepancy may be caused by different factors such as (1) imperfect spectral power balancing of the broadband signal and idler light within the LO and (2) the broad spectra of signal and idler light extending to the part of SFWM phase-matching spectrum where squeezing is considerably decreased.

\section{Discussion}
\rmd{mention sfwm flexibility before}

\begin{figure}
	\centering
	\includegraphics[width=0.8\columnwidth]{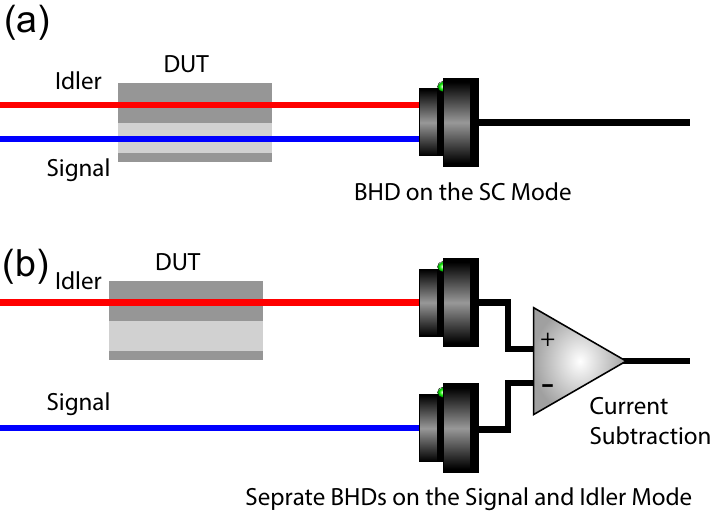}
	\caption{Comparison between (a) BHD of SC mode squeezing after interaction with a device under test (DUT) and (b) an equivalent implementation of two independent BHDs on signal and idler mode, respectively. \label{compare_multiplexed_independent_homodyne}}
\end{figure}
The 7.50 dB squeezing measured on a bichromatic mode demonstrates the capability of generating strong squeezing on SC modes. This result surpasses previous implementations of fully guided-wave squeezed light sources based on \(\chi\od{2}\) and \(\chi\od{3}\) nonlinearity. The squeezing limiting factors are mainly the insertion losses of fiber-optical components (FBGs, the BHD coupler), the non-ideal detection quantum efficiency, and the existence of parasitic Raman noise. The existence of 4.38 dB and 0.88 dB squeezing are confirmed on partially coherent and chaotic SC modes (that respectively correspond to using randomly modulated laser light and filtered ASE light as the LO). This confirms the ability to generate squeezing on SC modes with flexible time-frequency properties. The squeezing on SC modes demonstrated can be considered as a generalization of the bichromatic LO technique \cite{marino2007bichromatic,embrey2016bichromatic,yang2021squeezed,shi2020observation} that allows measuring squeezing far away from the pump frequency beyond the limit of photon detection bandwidth. This helps avoid degradation of squeezing due to pump-induced GAWBS noise \cite{shelby1986broad,shelby1985guided}, a long-standing obstacle for achieving strong squeezing with self-phase modulation in fiber. The impact of GAWBS can also be mitigated by using short pump pulses\cite{rosenbluh1991squeezed,friberg1996observation,bergman1991squeezing,bergman1994squeezing} and measuring polarization squeezing\cite{corney2008simulations,dong2008experimental}, but at the cost of requiring specialized source and detection system.\\

The SC mode squeezing source is implemented entirely with off-the-shelf fiber-optical components and can be packaged into a compact, modular form factor, with thermal and mechanical stability ensured by a high-speed optical phase-locked loop. Such a form factor, together with the flexible time-frequency properties and substantial magnitude of the squeezing, can serve as a useful modality for a broad range of squeezed light based applications. Specifically, for existing sensing systems that can be implemented in an all-fiber platform, such as a fiber-optical gyroscope\cite{haus1991fiber,eberle2010quantum,fink2019entanglement,grace2020quantum}, the substantial squeezing generated in the fiber source can be efficiently utilized due to the negligible coupling loss between the source and the sensing system via fusion splicing. However, it is worth noting that phase stabilization for squeezing enhanced sensing systems is more stringent: typically an additional classical coherent light source is required at the input of the sensing system, in addition to the squeezed vacuum. In such cases, the phases between the coherent light source, the squeezed vacuum and the LO need to be mutually locked. On this regard, we anticipate that the all-fiber phase-locked loop we develop here can be adapted for such sensing systems, with sufficient locking accuracy and minimal insertion loss. \\

As has been discussed, with heralding BHD of the idler mode, SC mode squeezing can be converted to single-mode squeezing on the arbitrarily shaped (in the time-frequency domain) signal mode as long as it is within the SFWM phase-matching bandwidth and the constant intensity portion of the pump pulse's temporal span. The heralding BHD can be done by physically separating the signal and idler mode of the SFWM output (i.e., with spectral filters), but at the cost of additional insertion loss for the squeezed light. Alternatively, the BHD of the SC can also be interpreted as independent BHDs (the heralding BHD of the idler mode and the BHD of the heralded signal mode) multiplexed on the same physical detector, and their respective photocurrents subtract each other. Inspired by this multiplexing interpretation, one can perform squeezing enhanced sensing in arbitrarily shaped signal mode by injecting the SC mode into the device under test (Fig. \ref{compare_multiplexed_independent_homodyne}(a)). Then as long as the impact of the device on the idler mode is known apriori, the BHD of the throughput SC mode can also be regarded as a simultaneous independent measurement on the (heralding) idler mode and the (heralded) arbitrarily shaped signal mode (Fig. \ref{compare_multiplexed_independent_homodyne}(b)).\\

It is worth noting that the flexibility of measuring squeezing in arbitrary SC modes is a result of the redundancy of broadband SFWM processes: a large number of close-to-degenerate squeezed SC modes exist within the SFWM output and BHD selectively measures one linear combination of them. Other unmeasured modes will contribute to extra shot noise in BHD, but such impact can be arbitrarily reduced using a strong LO light, thanks to the modal selective nature of BHD. Also, for squeezing enhanced sensing applications that use squeezed coherent light with a strong coherent component, broadband noise power in other unmeasured modes can be considered negligible. Broadband noise in other modes can also be completely removed through a modal selective quantum frequency conversion process that matches the profile of the signal and idler mode at the cost of insertion loss.\\

Despite that using standard telecom fiber to generate squeezing provides the unique benefit of low propagation and coupling loss, the underlying nonlinear process also poses certain constraints on the squeezing performance. First, the ``arbitrariness'' of an SC mode (with substantial squeezing) is limited by the SFWM phase-matching bandwidth in single-mode fiber. This is mainly due to (1) the use of the 1.5 km long fiber for sufficient nonlinear interaction and (2) the non-zero dispersion of single-mode fiber around 1550 nm. A rough estimate of the total number of squeezed SC modes generated by each pump pulse can be obtained from the time-bandwidth product of the pump pulse duration and SFWM phase-matching bandwidth. Second, it is also worth noting that the current system is not capable of directly (without heralding detection on the signal or idler mode) generating single-frequency squeezed light. To do so would require restricting the squeezing spectrum to the proximity of the pump frequency where dominant GAWBS noise exists. Third, the existence of spontaneous Raman noise will lead to diminishing scaling of measurable squeezing as pump power increases. Although cryogenic cooling can reduce the temperature-dependent Raman noise, it at the cost of inducing additional fiber microbending loss\cite{hass1969temperature,guo2012all}. \\

	Some of these aforementioned limitations of using telecom fiber can be circumvented by using other fiber structures or optical materials instead. For example, by using fiber with reduced dispersion or shorter length (with higher pump peak power), the phase-matching constraint can be less dominant, allowing larger phase-matching bandwidth\cite{guo2016generation,andrianov2022optimizing}. Besides silica, fiber-based on other materials may further extend the wavelength range of squeezing\cite{andrianov2024polarization,sorokin2022towards}. Notably, the possibility of using broadband squeezed light generated in fiber for sensing has been recently demonstrated for dual-comb spectroscopy\cite{herman2025squeezed}. 
The experimental setup for generating squeezing in SC modes can also be easily adapted for \(\chi\od{2}\) nonlinear medium, except that pump light with higher frequency is used. In comparison to SFWM, spontaneous parametric down-conversion is naturally free of Raman and GAWBS within the frequency range of interest, and the short length of \(\chi\od{2}\) waveguide allows over several THz of phase-matching bandwidth\cite{kashiwazaki2020continuous}. In addition, the high material nonlinearity and confinement of \(\chi\od{2}\) waveguide, especially in the form of thin film lithium niobate \cite{chen2022ultra}, may provide a viable route to system-level on-chip integration of generation and utilization of SC mode squeezing.  Nevertheless, as compared to fiber-based sources, the photodetection efficiency for waveguide sources still needs to be improved with low loss output coupling\cite{hu2021high} or high performance integrated photodiodes\cite{tasker2021silicon}, in order for it to generated comparable measurable squeezing.  

In conclusion, we propose and experimentally demonstrate a self-conjugated-mode squeezing system based on SFWM in fiber. The flexibility provided by time-frequency entanglement, for the first time, enables squeezing measurement on arbitrary self-conjugated time-frequency modes including those defined by noise-modulated laser (4.38 dB) and completely incoherent chaotic ASE light (0.88 dB). This can help pave the way towards generating squeezing on an arbitrary time-frequency mode. Self-conjugated mode squeezing also helps avoid the effect of pump-induced GAWBS noise in fiber, achieving higher measurable squeezing (7.50\(\pm\)0.12 dB) than previous implementations of guided wave squeezed light sources. The setup can be implemented in a compact, modular form factor exclusively with off-the-shelf standard telecommunication components that feature exceptionally low coupling loss to external fiber-based systems. We expect that such flexibility and efficiency in generating and measuring high squeezing can benefit a broad range of quantum information and sensing-metrology applications.

\section{Methods}
\subsection{Derivation of Self-conjugated Mode Squeezing Without the Broadband Phase-matching Approximation}
When the broadband phase-matching approximation is not perfectly satisfied (i.e. \(\xi(\omg) = |\xi(\omg)|\exp(i\phi)\) is complex and non-uniform over \(\omega\)), the SFWM evolution Eq.\eqref{USFWM0} states that single mode squeezing exists on bichromatic supermodes \(a_\text{even}(\omg),a_\text{odd}(\omg)\) (for different \(\omg>\omg_0\)):         
\begin{gather}
	a_\text{even}(\omg) = \exp\left(-i\frac{\phi(\omg)}{2}\right)\frac{a(\omg)+a(2\omg_0-\omg)}{\sqrt{2}} \\
	a_\text{odd}(\omg) =  \exp\left(-i\frac{\phi(\omg)}{2}+\frac{i\pi}{2}\right)\frac{a(\omg)-a(2\omg_0-\omg)}{\sqrt{2}} 
\end{gather}
with squeezing parameter \(\xi_\text{even}(\omg) = \xi_\text{odd}(\omg) =-|\xi(\omg)|\). Then consider an SC mode that is defined as an arbitrary real-factored linear combination of \(a_\text{even}(\omg),a_\text{odd}(\omg)\). It can always be expressed as: 
\begin{gather}
	a_\text{conj} =\\
	\sqrt{2}\int_{\omg_0}^{+\infty}\left(\text{Re}(f_\text{conj}(\omg))a_\text{even}(\omg)+\text{Im}(f_\text{conj}(\omg))a_\text{odd}(\omg)\right)d\omg\\
	= \int_{-\infty}^{+\infty}\exp\left(-i\frac{\phi(\omg)}{2}\right)f_\text{conj}(\omg)a(\omg)d\omg\\
\end{gather}
where \(f_\text{conj}(\omg)\) is arbitrary SC spectral amplitude satisfying Eq.\eqref{SCDEF}.
Note that since different even and odd modes have non-degenerate (frequency dependent) squeezing parameter and their linear mixing (with real coefficients) constitute the SC mode, the squeezing on the SC mode is given by the spectral average of squeezed shot noise of every even and odd mode:
\begin{gather}
	\avg{\Delta^2 (U\dg X_\text{conj}U )}\nonumber\\
	A= \frac{1}{2}\int_{-\infty}^{+\infty} |f_\text{conj}(\omg)|^2 \exp(-2|\xi(\omg)|)d\omg
\end{gather}
where \(X_\text{conj} = (a_\text{conj}+a\dg_\text{conj})/\sqrt{2}\) and the \(1/2\) factor accounts for the input vacuum fluctuation \(\avg{\Delta^2 X_\text{conj}}=1/2\).

\subsection{Components of the Experimental Setup}
The pump (Rio\rg ORION at 1549.96 nm) and signal (Santec\rg TSL-550 at 1549.65 nm) laser are modulated using 5 ns (Siglent\rg SDG6022X) and 3 ns (Highland Technology\rg T130, synced) pulses at 18 MHz and high extinction electro-optical amplitude modulators (Keyang Photonics\rg and Optilab\rg). The ASE of the pump erbium-doped fiber amplifier (Amonics\rg) is filtered using a combination of fiber Bragg grating and optical circulator (not shown). To balance the power of the signal and idler component of the LO light, the center frequency and bandwidth of a tunable band-pass filter (Alnair labs\rg bvf200) such that the signal component is slightly attenuated. This filter can be replaced with a temperature-controlled fiber Bragg grating filter. To ensure equal splitting of the LO light on the BHD coupler (Newport\rg, F-CPL-1550-N), the LO polarization is fine-tuned by an electrically variable fiber polarization controller (OZ Optics\rg, not shown). The resolution and video bandwidth of all electrical spectrum measurements are setup to 100 kHz and 10Hz. \\

\subsection{SFWM Phase Matching and Quantum Noise Spectrum Measurement}
\begin{figure}
	\centering
	\includegraphics[width=0.8\columnwidth]{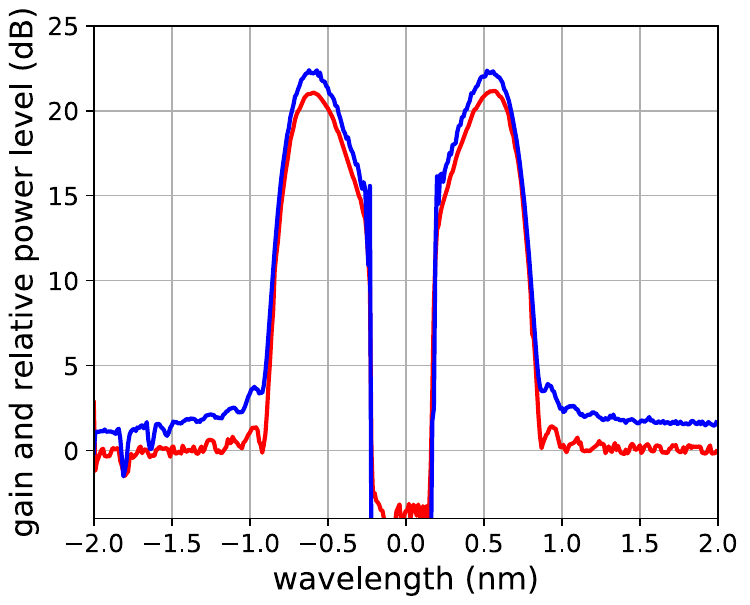}
	\caption{The PIA gain (red) and quantum (blue) spectrum of SFWM with 219 mW average pump power. The wavelength range around the pump frequency is not measurable because of pump filtering FBGs. \label{classical_spectra}}
\end{figure}

The SFWM phase-matching spectrum light is characterized in terms of the phase insensitive amplification (PIA) gain \(G_\text{PIA}(\omg)\): injecting signal seed light with power \(P_\text{seed}\) at frequency \(\omg\) together with pump light into the SFWM arm and measure the total output signal and idler power \(P_s, P_i\). The PIA gain \(G_\text{PIA}(\omg)\) relates to joint spectral amplitude \(\xi(\omg)\) via:
\begin{gather}
	G_\text{PIA}(\omg) = (P_s+P_i)/P_\text{seed} = 2\sinh^2(|\xi(\omg)|)+1
\end{gather}
The quantum noise spectrum of the SFWM output is measured through balanced heterodyne detection by mixing the SFWM output with a frequency-swept laser source that has the same pulse shape as the signal light. The measured quantum noise level is normalized with respect to the shot noise level when the SFWM is turned off. The measured quantum noise and gain spectra are shown in Fig.\ref{classical_spectra}.

\bibliographystyle{unsrt}
\bibliography{references.bib}
\section*{Data Availability}
The source data for the figures is available in (PLACEHOLDER FOR FIGURESHARE LINK).
\section*{Acknowledgements}
The project has been supported by NSERC an IDEaS programs.
\section*{Author contributions}
H.L. and A.H. conceived the idea. H.L. and M.L.I. designed the experiment. H.L. and M.L.I. and N.M.  conducted the experiment. H.L. and A.H. wrote the manuscript. All authors discussed the result and commented on the manuscript.
\section*{Competing interests}
The authors declare no competing interests.
\end{document}